\documentclass[showpacs,preprintnumbers,amsmath,amssymb]{revtex4}
\usepackage{amsmath,amssymb,amsfonts,latexsym,times}
\usepackage{citesort}
\usepackage{graphicx}
\usepackage{dcolumn}
\usepackage{bm}
\begin{document}
\preprint{ASC-LMU 03/06}

\title{Grand Unification in the $SO(10)$ Model}
\author{Alp Deniz \"Ozer and Harald Fritzsch }
\affiliation{Arnold Sommerfeld Center, University of Munich
 Theresienstr.37, DE-80333 M\"unchen }

\date{\today}

\begin{abstract}
Extrapolating the coupling strengths to very high energies, one
finds that they do not converge to a single coupling constant, as
expected in the simplest gauge theory of Grand Unification, the
$SU(5)$ theory.

We find that the coupling constants do converge, provided that two
new intermediate energy scales are introduced, the energies where
the group $SU(4)$, containing the color group, and the
right-isospin gauge group are spontaneously broken down.

\end{abstract}


\maketitle

\subsection{Introduction}

The Standard Model describes the electromagnetic, weak and strong
gauge interactions  and is based on the group $G_{sm} \equiv
SU(3)_C \times SU(2)_L \times U(1)_Y$
~\cite{Salam:1964ry,Weinberg:1967tq,Salam:1968rm,Glashow:1961tr,Fritzsch:2002jv,Fritzsch:1973pi,Weinberg:1973un,Glashow:1970gm,'tHooft:1972fi}.
Due to vacuum polarization effects the coupling strengths of the
three gauge interactions depend on the momenta of the
interaction~\cite{Aitchison}. The renormalization group formalism
gives rise to the so-called running coupling
constants~\cite{Georgi:1974yf}. Consequently, the coupling
constants are functions of the energy $Q$. At a few GeV, the
strengths of the weak and electromagnetic interactions are equal
to $\alpha_W \cong 1/30$ and $\alpha \cong 1/137$, whereas the
strong interaction strength varies in the range $2-100$ GeV as
$\alpha_s \approx 0.15 - 0.1 $~\cite{groom:2000}. These three
coupling strengths differ from each other in value. However the
weak and the electromagnetic gauge interactions become gradually
comparable at the Fermi energy scale $\approx 246$ GeV. As the
electroweak gauge symmetry $ SU(2)_L \times U(1)_Y$ is
spontaneously broken
down~\cite{Goldstone:1961eq,Goldstone:1962es,Higgs:1964pj,Higgs:1964ia,Higgs:1966ev}
at $(\sqrt{2} \, G_F)^{-1/2} \approx 246 $ GeV through a Higgs
doublet whose electrically neutral component acquires a non-zero
vacuum expectation value, the electrically neutral gauge fields
undergo a mixing so that the resulting massless gauge field
couples to the electromagnetic current. Thereby the
electromagnetic and the weak coupling strengths become related
through a mixing angle~\cite{Weinberg:1971nd}. We have
\begin{equation}
    \sqrt{\alpha} = \sqrt{\alpha_W} \sin \theta_W = \sqrt{\alpha_Y} \cos
    \theta_W
\end{equation}
Hence the mixing angle describes a {\it partial unification} of the
two distinct interactions~\cite{Weinberg:1979pi}.
The non-Abelian nature of the weak and the strong interactions give
rise to self interactions among the gauge
fields~\cite{Politzer:1973fx,Gross:1973id,Politzer:1974fr} in which
the self-coupling vertices cause an {\it anti-screening} effect.
The couplings strengths decrease with increasing $Q$. In
the contrary, the coupling strength of an Abelian interaction
increases with increasing energy. One finds
\begin{equation}\label{ren}
 \frac{1}{g^{2}_i(\mu)} = \frac{1}{g^{2}_{i}(Q)} + 2b_i \ln
 \frac{Q}{\mu}  + \text{higher orders}
\end{equation}
where $g_i(\mu)$ and $g_i(Q)$ are the measured strength of $g_i$ at
the energy scale $\mu$ and $Q$ respectively. Formally the
renormalization of any of the gauge coupling $g_i$ depends  on the
dimension of the unitary gauge group to which the coupling is
assigned. This property is contained in the $b_i$ functions which
get contributions from gauge bosons, fermion loops and scalar
bosons~\cite{Jones:1974mm}. If only  the gauge bosons and fermion
loop contributions to vacuum polarization are taken into account one
obtains
\begin{equation}
  b_N =\frac{1}{(4\pi)^2} \left[ -\frac{11}{3}N + \frac{4}{3}n_g\right]
\end{equation}
where $N > 1$ is related with $SU(N)$ and $n_g = 3$ indicates the
number of fermion generations. Using the equations given above,
the values of $\alpha , \alpha_W , \alpha_Y$ at the {\it Fermi
scale} can be extrapolated towards higher energy scales. They
converge with increasing energy $Q$ and get very close in the
energy interval $10^{13}$ GeV to $10^{17}$ GeV. But a common
intersection of the three coupling strengths does not occur.
However the convergence of the coupling strengths suggests a
unification in which all gauge interactions posses the same
strength at some characteristic energy scale.

One possibility is to embed $G_{SM}$ into a gauge group that has a
single gauge coupling. The simplest possible choice would be the
$SU(5)$~\cite{Georgi:1974sy,Georgi:1974my} which has the same rank
as $G_{sm}$. Besides the gluons, the $W$, $Z$ and $\gamma$ there
are lepto-quark gauge bosons in the theory. (they can transform
quarks into leptons). Consequently they can mediate nucleon
decay~\cite{Gell-Mann:1976pg,Machacek:1979tx}, giving rise to
physics beyond the standard model~\cite{Weinberg:1979sa}. We
denote them with $Y_\alpha,Y^\prime_\alpha$, where $\alpha$
denotes $SU(3)$-color. Their charges are shown in Table.
\ref{qnbosons}. The lifetime of the proton to decay into a $\pi^0
\, e^+$ can be estimated as
\begin{equation}
    \tau_p \propto \frac{M^4_G}{\alpha^2_G m^5_p}
\end{equation}
where $m_p$ is the proton mass, $M_G$ is the lepto-quark gauge
boson mass and $\alpha_G$ is the coupling strength at
$M_G$~\cite{Goldman:1979ij,Ross:1980gh}. The proton life time is
sensitive to the gauge boson masses that mediate the decay
process~\cite{Ross:1980gh}\cite{Goldman:1980ah}\cite{Goldhaber:1980dn}\cite{Langacker:1992qb}.
A grand unification mass scale should not lie lower than $M_G \sim
10^{15}$ GeV. Immediately after the spontaneous breakdown of the
$SU(5)$ symmetry at $M_G$ through a singlet, the residual symmetry
group will be equal to that of the Standard Model gauge group. By
construction, the relative gauge couplings of the residual gauge
symmetry are equal at this energy scale. But as we run them down,
they depart from their known values at the Fermi scale.
Consequently they can not reproduce the experimental numbers. A
contribution of a Higgs doublet has also no considerable
influence. Thus no convergence takes place.

This problem disappears, however, if at energies above $1$ TeV
supersymmetry appears. In this case the coupling constants do
converge at the energy about $10^{16}$ GeV. In such a theory a new
energy scale arises, the energy where supersymmetry is realized.

In the {\it minimal} $SU(5)$ theory wherein the Higgs sector can
be adjusted to exhibit coupling unification, the grand unification
mass scale roughly takes place at $3.1 \times 10^{14\pm 0.3}$ GeV
which is relatively low~\cite{kounnas}. Consequently the {\it
minimal} $SU(5)$ theory predicts the proton to decay in $2 \times
10^{29\pm 1.7}$ years~\cite{kounnas} which is faster than the
recently measured lower bound~\cite{groom:2000} and therefore the
theory is in serious trouble~\cite{Langacker:1980js}.

We shall investigate here a further possibility of coupling
unification by  embedding  $G_{sm}$ into the $SO(10)$ gauge
group~\cite{Fritzsch:1974nn}.

\subsection{$SO(10)$ gauge interactions}

The Lagrangian of the massless spinor field describing the $SO(10)$
gauge interactions can be stated as
\begin{equation}\label{spinor-L}
\begin{split}
\mathcal{L } & =   \bar{\Psi} \ i \  \gamma^{\mu} \ D_{\mu} \   \Psi
- \frac{1}{4} \, {F^{\mu \nu}}^{\ ab} {F_{\mu \nu}}^{ab} \\ & =
\underbrace{\bar{\Psi} \ i \ \gamma^{\mu} \
\partial_{\mu} \Psi }_{\text{K.E. of $\Psi$
}}- \ \underbrace{\frac{g}{\sqrt{2}} \  \bar{\Psi} \, \left( \,
W_{\mu} \cdot \Sigma \, \right) \, \Psi}_{\text{Interactions}} -
\underbrace{\frac{1}{4} \, {F^{\mu \nu}} \cdot {F_{\mu
\nu}}}_{\text{K.E. of W's}}
\end{split}
\end{equation}
where $\Psi$ is the family spinor
that contains the fermions of a single generation and $D_\mu$ is
the gauge covariant derivative~\cite{ozer}. It is defined as
\begin{equation}\label{cov-der}
D_{\mu} = \partial_{\mu} + i \ \frac{g}{\sqrt{2}} \ W_{\mu}  \cdot
\Sigma
\end{equation}
Here $W^{ab}_\mu$  are real valued $N (N-1) /2$ vector gauge fields
with $a,b=1,\dots,N$; $N=10$, $\Sigma_{ab}$ are the representation
matrices and $g$ is the single coupling strength. The above inner
product implies a sum over the group indices $ab$. Furthermore
$D_\mu$ is a matrix in the space of group indices, and $\Sigma_{ab}
= -\Sigma_{ba}$. The number of independent physical gauge fields is
determined by the degrees of freedom possessed  by the unification
gauge group.

For $SO(10)$, we have 45 gauge fields. The second term
appearing in the covariant derivative is usually known as the gauge
term matrix. The physical content of the gauge term matrix can be
highlighted by using an explicit matrix representation of the
$\Sigma$'s. Several examples are given in ref.~\cite{ozer}.
Since the $\Sigma$'s transform the family spinor like $\Psi_{a}
\rightarrow \left({ e^{-i \ \Sigma \cdot  \omega} } \right)_{ab}
\Psi_{b}$ $=$ $U_{ab} \ \Psi_{b}$, one should consider the unitary
(or spinorial) representation of $SO(10)$. The $\Sigma_{ab}$ basis
of the spinorial $SO(10)$ representation can be generated through a
$\Gamma$ basis which  satisfies the {\it Clifford
Algebra}~\cite{Bauer:1952}. We have
\begin{equation}
\Sigma_{ab}=\frac{i}{4}[\Gamma_{a},\Gamma_{b}] \ \ , \ \ \ \{
\Gamma_{a} ,  \Gamma_{b}  \} = 2 \ \delta_{a b} \ 1\hspace{-2.3mm}1
\end{equation}
Here curly brackets denote anticommutation. $1\hspace{-2.3mm}1$ is
a unit matrix with size $32 \times 32$. The $\Sigma$'s satisfy the
{\it Lie Algebra of } $SO(10)$. We have
\begin{equation}
{[\Sigma_{ab} ,\Sigma_{cd} ] = \, i \left( \, \delta_{ad} \
\Sigma_{bc} + \delta_{bc} \ \Sigma_{ad} -\delta_{ac} \ \Sigma_{bd}
-\delta_{bd} \ \Sigma_{ac} \,\right)}
\end{equation}
It follows that a maximal sub-group of $SO(10)$ is the $SO(6) \times SO(4)$ where
$SO(6)$ is spanned by the indices $a,b = (1,\dots,6) $ and $SO(4)$
is spanned by the indices $a,b = (7,\dots,10) $. $SO(4)$ is
isomorphic to $SU(2) \times SU(2)$ which defines two mutually commuting
isospin-triplet gauge fields. These are the $W^{0,\pm}_L$ and $W^{0,\pm}_R$ fields
which couple to left- and right-handed currents respectively.

$SO(6)$ is isomorphic to $SU(4)$ and contains  $SU(3)_c \times
U(1)_{B-L}$ as a subgroup which defines the $8$ Gluons $G_i$ and a
singlet $X_{B-L}$ field. In the coset $SU(4)/SU(3)_c$, we have a
sextet field which decomposes with respect to the $U(1)_{B-L}$ group
into two charge conjugated color triplets $3$ and $\bar{3}$. These
are the $X_\alpha$ lepto-quark fields where $\alpha = (r,g,b) $
denotes $SU(3)$-color~\cite{Pati:1974yy,Pati:1977jh}. On the other
side the coset $SO(10) / SO(6) \times SO(4)$ contains $24$
leptoquark gauge fields denoted with $A_\alpha,
A^\prime_\alpha,Y_\alpha,Y^\prime_\alpha$ which decompose into a
bi-doublet sextet. All these $45$ gauge fields and their various
charges are summarized in Table.~\ref{qnbosons}.

\begin{table}[htb]
\begin{center}
\begin{tabular}{ccccccccccccc}
  \hline \hline
  \vspace{-2mm} &&&&&&&&&&&& \\
                & $Q$  &$B-L$ &$I_{3R}$&$I_{3L}$&$Y $  &    \ \   &     & $Q$  &$B-L$ &$I_{3R}$&$I_{3L}$&$Y $   \\
  \vspace{-2mm} &&&&&&&&&&&& \\
  \hline
  \vspace{-2mm} &&&&&&&&&&&& \\
   $G_i$      &0  &0  &0        &0        &0             & \ \ &$W^{+}_{L}$&+1 &0  &0        &+1       &0  \\
    \vspace{-2mm} &&&&&&&&&&&&    \\
   $X_{B-L}$  &0  &0  &0        &0        &0             & \ \ &$W^{0}_{L}$&0  &0  &0        &0        &0   \\
  \vspace{-2mm} &&&&&&&&&&&&    \\
   $X_{\alpha}$    &2/3&4/3&0        &0        &2/3      & \ \ &$W^{-}_{L}$&-1 &0  &0        &-1       &0    \\
   \vspace{-1mm} &&&&&&&&&&&&     \\
  $A_{\alpha}$       & 2/3&-2/3&+1/2    &+1/2     &1/3   & \ \ &$W^{+}_{R}$&+1 &0  &+1       &0        &+1 \\
  \vspace{-2mm} &&&&&&&&&&&&    \\
  $A^\prime_{\alpha}$&-1/3&-2/3&+1/2     &-1/2     &1/3   & \ \ &$W^{0}_{R}$&0  &0  &0        &0        &0   \\
  \vspace{-2mm} &&&&&&&&&&&&    \\
  $Y_{\alpha}$       &-1/3&-2/3&-1/2     &+1/2     &-5/3  & \ \ &$W^{-}_{R}$&-1 &0  &-1       &0        &-1  \\
  \vspace{-2mm} &&&&&&&&&&&&    \\
  $Y^\prime_{\alpha}$&-4/3&-2/3&-1/2     &-1/2     &-5/3   & \ \    &&&&&& \\
\hline \hline
\end{tabular}
\caption[The 45 gauge bosons mediating $SO(10)$ gauge
interactions]{Charges of the 45 Gauge Bosons}\label{qnbosons}
\end{center}
\end{table}
In contrast to the electroweak theory, it is a rather
sophisticated   task to illustrate the content of the $SO(10)$
gauge term through a matrix. The gauge term reads
\begin{equation}
\begin{split}
 +i  \frac{g}{\sqrt{2}} W_{ab} \hspace{1mm} \Sigma_{ab}
  & = + i\, g \, \sqrt{2} \,   \left\{ G \cdot U_{G} +  \left( \ X \cdot U_{X} + h.c.
  \right)+
  + \sqrt{\frac{3}{2}} \frac{X_{B-L}}{\sqrt{2}} \, \frac{U_{B-L}}{2}
  + W^{\pm}_L L_{\pm}+  \frac{W^{3}_L}{\sqrt{2}} L_{3} \right. \\
  & \left. + W^{\pm}_R R_{\pm} + \frac{W^{3}_R}{\sqrt{2}} R_{3}+(D_{A} \cdot A+D_{A^\prime}
  \cdot A^\prime+D_{Y} \cdot Y +D_{Y^\prime} \cdot Y^\prime + h.c. ) \right\} \\
  & = +i \frac{g}{\sqrt{2}} \left[
 \begin{array}{cc}
  \Lambda_{11}     &   \Lambda_{12} \\
 \Lambda_{21}     &   \Lambda_{22}  \\
\end{array}\right] \\
\end{split}
\end{equation}
The generators multiplying the (neutral) singlet fields,
$W^3_L,W^3_R$ and $X_{B-L}$ add up to the electric-charge
eigenvalue operator $Q $ = $L_3+R_3$ $+U_{B-L}/2$ where $L_3$ and
$R_3$ are the left and right-isospin eigenvalue operators and
$U_{B-L}$ is the $B-L$ eigenvalue operator. The entries
$\Lambda_{11},\Lambda_{12},\Lambda_{21}$ and $\Lambda_{22}$ are
matrices each of size $16 \times 16$. They explicitly  depend on
the $\Gamma$ basis as studied in~\cite{ozer}. The interaction term
given in eq.~(\ref{spinor-L}) generates the currents coupling to
the fermions in the family spinor $\Psi$. We have
\begin{equation}
\begin{split}
 +i \frac{g}{\sqrt{2}} \bar{\Psi} \gamma_{\mu}
 \Sigma \cdot {W^{\mu}}
 \Psi & = + i \,   g \, \sqrt{2} \,  \left\{ \left( J^{A}_{\mu} \cdot {A^{ \mu}} +J^{A^\prime}_{\mu} \cdot  {A^{ \prime \mu}} +
 J^{Y}_{\mu}  \cdot {Y^{ \mu}}+J^{Y^\prime}_{\mu}  \cdot {Y^{ \prime \mu}} + h.c.
 \right)+
 \right. \\
 &
  \left( J^{X}_{\mu } \cdot {X^{ \mu}} + h.c. \right)
 + J^{\mu }_{B-L} \frac{X_{\mu \, B-L}}{\sqrt{2}}
 + J_G \cdot G
 + \left.  J^{\pm}_R W^{\pm}_R
 +  J^{3}_R \frac{W^{3}_R}{\sqrt{2}} + J^{\pm}_L W^{\pm}_L
 +  J^{3}_L \frac{W^{3}_L}{\sqrt{2}}  \right\} \\
 \end{split}
\end{equation}
The  currents $J^{A}_{\mu}, J^{A^\prime}_{\mu} $ etc. and the
gauge fields are given in the flavor basis. Depending on the
content of the Higgs sector, the gauge fields and thereby the
currents can undergo some complex mixing. A detailed analysis of
these mixing angles and phases which are possible sources of
$CP$-violation will not be studied in this paper.

\subsection{Renormalization of $\alpha$'s in $SO(10)$}

\begin{figure}[htb]
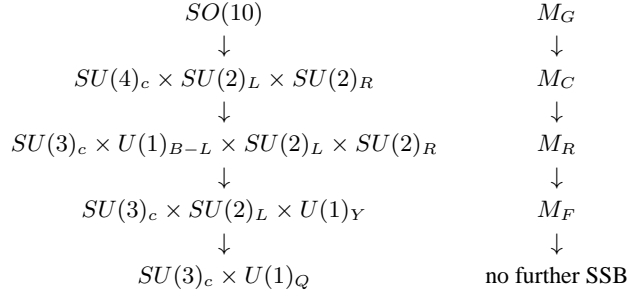

\begin{eqnarray*}
\begin{array}{ccc}
 & SO(10)     &       \\
  \vspace{-3mm} &            &       \\
 & \downarrow &       \\
  \vspace{-3mm} &            &       \\
 & SU(4)_{c} \times SU(2)_{L} \times SU(2)_{R} &   \\
  \vspace{-3mm} &            &       \\
 & \downarrow &       \\
  \vspace{-3mm} &            &       \\
 & SU(3)_{c} \times  U(1)_{B-L} \times SU(2)_{L} \times SU(2)_{R}   &      \\
  \vspace{-3mm} &            &       \\
 & \downarrow &       \\
  \vspace{-3mm} &            &       \\
 & SU(3)_{c} \times  SU(2)_{L} \times U(1)_{Y}  &  \\
  \vspace{-3mm} &            &       \\
 & \downarrow &       \\
  \vspace{-3mm} &            &       \\
 & SU(3)_{c} \times  U(1)_{Q}   &    \\
  \vspace{-3mm} &            &       \\
  \vspace{-3mm} &  &       \\
\end{array}
\begin{array}{ccc}
 & M_G     &       \\
  \vspace{-3mm} &            &       \\
 & \downarrow &       \\
  \vspace{-3mm} &            &       \\
 & M_C &   \\
  \vspace{-3mm} &            &       \\
 & \downarrow &       \\
  \vspace{-3mm} &            &       \\
 & M_R   &      \\
  \vspace{-3mm} &            &       \\
 & \downarrow &       \\
  \vspace{-3mm} &            &       \\
 & M_F &  \\
  \vspace{-3mm} &            &       \\
 & \downarrow &       \\
  \vspace{-3mm} &            &       \\
 & \text{no further SSB}   &    \\
  \vspace{-3mm} &            &       \\
  \vspace{-3mm} &  &       \\
\end{array}
\end{eqnarray*}
\caption[Backbone of symmetry descents in $SO(10)$]{
Descents in $SO(10)$ are on the left hand side and corresponding
intermediate mass scales are on the right hand side.
}\label{descents}
\end{figure}

At energies above the Grand Unification energy mass scale $M_G$
there is a single coupling constant $g$. The relative
coupling strengths $g_i$ assigned to the various subgroups of
$SO(10)$ are equal in value at the  grand unification mass scale
$M_G$ whose value is subject to further determination. We define
\begin{equation}\label{strengths}
\begin{split}
g & = \underbrace{\frac{g_L}{C_L}}_{g_1}=
\underbrace{\frac{g_R}{C_R}}_{g_2}
 = \underbrace{\frac{g_{B-L}}{C_{B-L}}}_{g_3}= \underbrace{\frac{g_Y}{C_Y}}_{g_4}
 = \underbrace{\frac{\sqrt{4 \pi
 \alpha_s}}{C_s}}_{g_5}  = \underbrace{\frac{\sqrt{4 \pi
 \alpha_{s^\prime}}}{C_{s^\prime}}}_{g_6}= \underbrace{\frac{e}{C_Q }}_{g_7}
\end{split}
\end{equation}
where $C$ are the Clebsch-Gordon coefficients. These coefficients
take the values $C_L=C_R=1$, $C_{B-L}=\sqrt{3/2}$,
$C_{Q}=\sqrt{3/8}$, $C_Y=\sqrt{3/5}$. They can be found by
rescaling the diagonal generators so that they  become eigenvalue
operators and produce charge quantization~\cite{ozer}. At $M_G$
the symmetry breaks down to $SU(4)_C \times SU(2)_L \times
SU(2)_R$. The corresponding coupling strengths are donated by
$\alpha_{s^\prime},g_L$ and $g_R$. At $M_C$ the symmetry breaks
down to $SU(3)_C \times SU(2)_L \times SU(2)_R \times U(1)_{B-L}$.
Finally at the energy $M_R$ the symmetry is broken down to
$SU(3)_C \times SU(2)_L \times SU(1)_Y$. The final breakdown to
$SU(3)_C \times U(1)_{Q}$ is similar to the symmetry breaking in
the standard model. These intermediate mass scales and symmetries
are illustrated  in Fig.~(\ref{descents}).
\noindent The evolution of the coupling strengths are given as
follows:
\begin{eqnarray}
 U(1)_Y \  ;  \frac{1}{g^2_4 (M_F)} & = & \frac{1}{g^2(M_G)} + 2 (\frac{2}{5} b_4 +
 \frac{3}{5} b^R_2)  \ln(\frac{M_G}{M_C}) + 2 (\frac{2}{5} b^C_1 +
 \frac{3}{5} b^R_2)\ln(\frac{M_C}{M_R})+ 2 b_1
 \ln(\frac{M_R}{M_F}) \nonumber \\
 SU(2)_L \  ;   \frac{1}{g^2_1(M_F)} & = & \frac{1}{g^2(M_G)} + 2 b^L_2 \ln(\frac{M_G}{M_F})  \nonumber \\
 SU(3)_C \  ;   \frac{1}{g^2_5(M_F)} & = & \frac{1}{g^2(M_G)} + 2 b_4   \ln(\frac{M_G}{M_C}) + 2 b_3   \ln(\frac{M_C}{M_F})\\
 U(1)_Q \ ;   \frac{3}{8}\frac{1}{e^2(Q)} & = & \frac{3}{8}\frac{1}{e^2(M_G)}
  + 2 ( \frac{2}{8}b_4 + \frac{3}{8}b^R_2
  + \frac{3}{8}b^L_2 ) \ln(\frac{M_G}{M_C})
    \nonumber \\ & & + 2 ( \frac{2}{8}b^C_1 + \frac{3}{8}b^R_2
  + \frac{3}{8}b^L_2 ) \ln(\frac{M_C}{M_R})
  + 2(\frac{5}{8} b_1 +  \frac{3}{8} b_2) \ln(\frac{M_R}{M_F})
   + 2 b^{em}_1   \ln(\frac{M_F}{Q})\nonumber
\end{eqnarray}
where the $g_i$ and  $b_i$ are defined as follows:
\begin{equation}\label{beta_all}
\begin{tabular}{llrcl}
   $U(1)_Q     \ \ ,  $ & $  b^{em}_1 \ \ , $ & $ e(Q)         $ & = & $ C_Q \, g_{7}(Q) = \sqrt{4\pi \alpha(Q)}       $ \\
 \vspace{-3mm} & & & & \\
   $SU(4)_C  \ \ ,  $ & $  b_4  \ \ ,    $ & $ \alpha_{s^\prime}(Q)        $ & = & $ \frac{g^{2}_6(Q)}{4\pi} $ \\
 \vspace{-3mm} & & & & \\
  \vspace{-3mm} & & & & \\
  $SU(3)_C    \ \ ,  $  & $  b_3  \ \ ,    $ & $ \alpha_{s}(Q)        $ & = & $ \frac{g^{2}_5(Q)}{4\pi} $  \\
  \vspace{-3mm} & & &  & \\
   $U(1)_Y     \ \ ,  $ & $  b_1 \ \ ,      $ & $ g_{Y}(Q)         $ & = & $ C_Y \, g_{4}(Q) $ \\
  \vspace{-3mm}   & & & & \\
  \vspace{-3mm} & & &  \\
   $U(1)_{B-L} \ \ ,  $ & $  b^c_1 \ \ ,   $ & $ g_{B-L}(Q)  $ & = & $  C_{B-L} \, g_{3}(Q) $ \\
 \vspace{-3mm} & & & & \\
 \vspace{-3mm}  & & & & \\
   $SU(2)_R    \ \ ,  $ & $  b^R_2 \ \ ,    $ & $ g_R(Q)                $ & = & $ C_R \, g_{2}(Q) $  \\
  \vspace{-3mm}   & & & & \\
 \vspace{-3mm} & & & & \\
   $SU(2)_L    \ \ ,  $ & $  b^L_2 \ \ ,   $ & $ g_L(Q)                $ & = & $ C_L \, g_{1}(Q) $ \\
\end{tabular}
\end{equation}
Note that in general we have $M_F < M_R < M_C < M_G$. From the
above equations the unknown scales $M_G$, $M_C$ and $M_R$ can be
found. A possibility is to use the the combination ;
$C_Y^{-2}/g^2_4+1/g^2_1-(C_Y^{-2}+1)/g^2_5$. It is free of the
unknown $g$ and the first two terms in this combination equal to
$1/e^2$. It yields;
\begin{equation}\label{1}
 \frac{1}{e^2}-\frac{8}{3}\frac{1}{g^2_5} = \mathcal{A}  \ln(\frac{M_G}{M_C})+\mathcal{B} \ln(\frac{M_C}{M_R})+ \mathcal{C} \ln(\frac{M_R}{M_F}) \\
\end{equation}
where the constants come out as
\begin{equation}\label{coef1}
 \begin{split}
 \mathcal{A} & = 2 (b^L_2 -  2 b_4 + b^R_2) \\
 \mathcal{B} & = 2 (\frac{2}{3} b^C_1 +   b^R_2- \frac{8}{3} b_3 +   b^L_2 ) \\
 \mathcal{C} & = 2 (\frac{5}{3} b_1  - \frac{8}{3}b_3 + b^L_2) \\
 \end{split}
\end{equation}
There are three unknowns in eq.~(\ref{1}), namely the scales $M_G$,
$M_C$ and $M_R$. Further relations are required. Another possibility
comes from the weak mixing angle contained by  the combination
$C^{-2}_Y/g^2_4-C^{-2}_Y/g^2_1$. We have
\begin{equation}
\frac{C^{-2}_Y}{g^2_4}-\frac{C^{-2}_Y}{g^2_1}=
\frac{1}{e^2}-(1+C^{-2}_Y)\frac{\sin^2\theta}{e^2}
\end{equation}
Using the expressions for  $g_4$ and $g_1$ and some reorganization
of the terms yield
\begin{equation}\label{2}
 \sin^2\theta =\frac{3}{8}- \frac{5 }{8} e^2\left(\mathcal{ D} \ln(\frac{M_G}{M_C})+ \mathcal{E }\ln(\frac{M_C}{M_R})+ \mathcal{F } \ln(\frac{M_R}{M_F})  \right)
\end{equation}
The constants $\mathcal{D}$, $\mathcal{E}$ and $\mathcal{F}$ in
terms of the beta functions are found as
\begin{equation}\label{coef2}
 \begin{split}
  \mathcal{D} & =  2 (\frac{2}{5} b_4 +
 \frac{3}{5} b^R_2 -   b^L_2  )  \\
  \mathcal{E} & =  2(\frac{2}{5} b^C_1 +  \frac{3}{5} b^R_2 -   b^L_2) \\
  \mathcal{F} & =  2 (b_1-    b^L_2) \\
 \end{split}
\end{equation}
We have  three unknown scales $M_G $, $M_C$ and $M_R$ in each of
the statements above. We make the assumption that $M_G$ and $M_C$
are close so that  $\ln \frac{M_G}{M_C} \gtrsim 0$. In terms of
this ratio, the eqs.~(\ref{1}) and~(\ref{2}) can be solved among
themselves for the three unknown scales (for details see Appendix
A).

\subsection{Mass Scales in $SO(10)$}
\label{m-scales}

\begin{figure}[h]
\begin{center}
\includegraphics[scale=1.0]{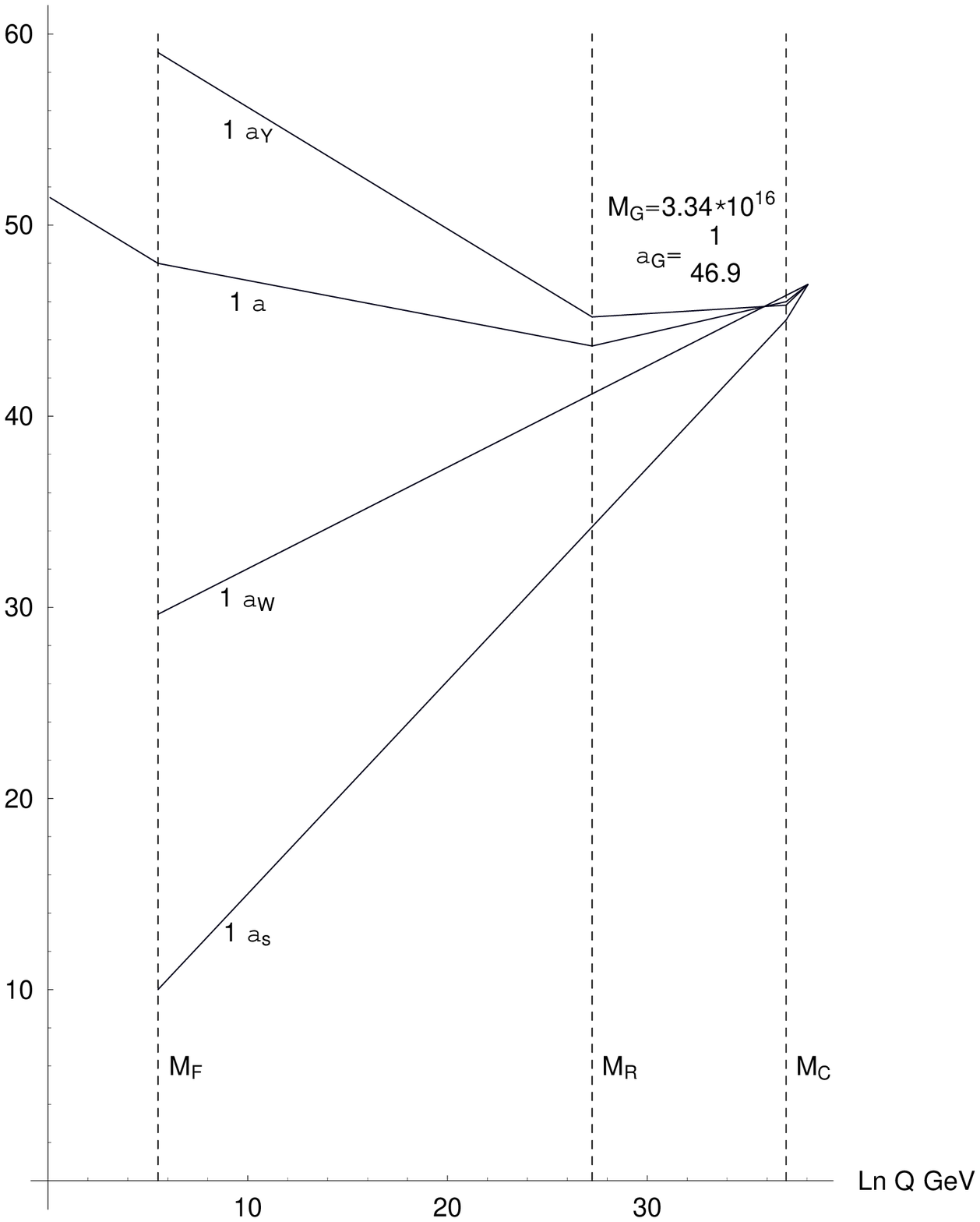}
\caption[The evolution of $1/\alpha(Q)$ with respect to $\ln
Q$.]{The evolution of $\alpha_s$, $\alpha_w$, $\alpha$, $\alpha_Y$,
with respect to $\ln Q$ where Q is in GeV.}\label{alpha}
\end{center}
\end{figure}

Let us estimate  the scales without any contribution of Higgs
scalars using eq.~(\ref{scales}). These values are summarized in
Table~\ref{ims}. It is found that $M_G$ rests between $10^{16}-
10^{17}$ GeV and $M_R$ between $10^{10}-10^{11}$ GeV for various
values of $\ln(M_G/M_C)$ and acceptable input values of the
electroweak parameters at $ M_F=(\sqrt{2} \, G_F)^{-1/2} \cong
246.22$ GeV where especially $\alpha_s^{-1}(M_F)=10$. For smaller
values of $\alpha_s^{-1}$, the $SO(10)$ model is no more
physically viable, because  $M_G$ moves inescapably towards the
Planck scale. The various values of $\alpha(M_G)$ are also given
in Table.~\ref{ims} for the respective values of the intermediate
mass scales where the fine structure constant $\alpha$ at $M_G$ is
obtained as
\begin{equation}
\begin{split}
  \frac{3}{8} \frac{(4 \pi)^{-1}}{\alpha(Q)}
  & = \frac{3}{8} \frac{(4 \pi)^{-1}}{{\alpha}(M_G)}
  + 2 ( \frac{2}{8}b_4 + \frac{3}{8}b^R_2 + \frac{3}{8}b^L_2 )
  \ln(\frac{M_G}{M_C})
   + 2 ( \frac{2}{8}b^C_1 + \frac{3}{8}b^R_2 + \frac{3}{8}b^L_2 ) \ln(\frac{M_C}{M_R})
  \\ &      + 2   (\frac{5}{8} b_1
  +  \frac{3}{8} b_2) \ln(\frac{M_R}{M_F}) + 2 b^{em}_1   \ln(\frac{M_F}{Q}) \ \ \
\end{split}
\end{equation}
This expression can be  solved for $\alpha(M_G)$ at $Q=M_F$. If we
compare the values of $\alpha^{-1}$ at $M_G$ and $M_F$, given in the
last six rows in Table.~\ref{ims}, it is observed that they are
very close. This does not happen in the effective $SU(3)_C \times
SU(2)_L \times U(1)_Y$ running of $\alpha^{-1}$. The difference is
caused by the existence of the $U(1)_{B-L} \times SU(2)_{R}$
symmetries above the $M_R$ scale. The beta functions $b^{em}_1$
in the interval $Q > M_R$ becomes negative. The
running of $\alpha,\alpha_W,\alpha_s,\alpha_Y$ are
sketched in Fig.~\ref{alpha}. As seen in
the figure $\alpha^{-1}$ reaches a minimum value of approximately
$117$ at the mass scale $M_R$, beyond this scale the electromagnetic
interactions become gradually weaker again.
\begin{table}[htb]
\begin{tabular}{ccccccccc}
 \hline \hline
   \vspace{-1mm}&&&&&&&& \\
 $\ln \frac{M_G}{M_C}$ &$\alpha^{-1}_s(M_F)$ & $M_G$                & $M_C$                   &  $M_R$                     & $\alpha^{-1}(M_G)$ &  $\sin^2\theta_W(M_F)$   &  $\alpha^{-1}(M_F)$  &  $\alpha^{-1}_G$ \\
  \vspace{-1mm} &&&&&&&& \\
  \hline
  \vspace{-1mm} &&&&&&&& \\
 $\ln1$      &     9       & $5.56 \times 10^{17}$ &  $5.56 \times 10^{17}$  &     $2.40 \times 10^{10}$      &    129.0   &         $0.2315$         &       128       &    48.4        \\
  \vspace{-2mm} &&&&&&&& \\
 $\ln2$      &     9       & $2.78 \times 10^{17}$ &  $1.39 \times 10^{17}$  &     $9.62 \times 10^{10}$      &    128.1   &         $0.2315$         &       128       &    48.0      \\
  \vspace{-2mm} &&&&&&&& \\
 $\ln3$      &     9       & $1.85 \times 10^{17}$ &  $6.16 \times 10^{16}$  &     $2.16 \times 10^{11}$      &    127.5   &         $0.2315$         &       128       &    47.8          \\
  \vspace{-2mm} &&&&&&&& \\
 \vspace{-2mm}  &&&&&&&& \\
 $\ln1$      &    10       & $1.00 \times 10^{17}$ &  $1.00 \times 10^{17}$  &     $7.54 \times 10^{10}$      &    126.6   &         $0.2315$         &       128       &     47.5          \\
  \vspace{-2mm} &&&&&&&& \\
 $\ln2$      &    10       & $5.02 \times 10^{16}$ &  $2.50 \times 10^{16}$  &     $3.02 \times 10^{11}$      &    125.6   &         $0.2315$         &       128       &     47.1           \\
  \vspace{-2mm} &&&&&&&& \\
 $\ln3$      &    10       & $3.34 \times 10^{16}$ &  $1.12 \times 10^{16}$  &     $6.78 \times 10^{11}$      &    125.1   &         $0.2315$         &       128       &     46.9          \\
  \vspace{-2mm} &&&&&&&& \\
 \hline \hline
\end{tabular}
\caption[Grand unification  and intermediate mass scales in
$SO(10)$] {Grand unification  and intermediate mass scales without
contribution of any Higgs scalars. $\alpha_G=g^2/4\pi$ is the
coupling strength at the grand unification mass scale $M_G$.}
\label{ims}
\end{table}
\subsection{Results and Discussion}
In this paper we have studied coupling unification in the
$SO(10)$-Theory. Unlike the $SU(5)$ theory this theory shows a
convergence of the coupling constants, if we adjust them at low
energies to the observed values. We have found that the coupling
constants do converge, provided that two new intermediate energy
scales are introduced, the energies where the 4-color gauge group
and the right-isospin gauge group are spontaneously broken down.
However two new energy scales have been introduced, the energy
where leptons and quarks start to become separated and $SU(4)$ is
broken down to the QCD gauge group, and the energy, where the
righthanded isospin is broken. The first energy scale, denoted by
$M_C$, is very close to $M_G$ and might be identical to $M_G$. The
second energy scale $M_R$ is around $10^{10-11}$ GeV. Similar
conclusions have also been reached in ref.~\cite{Lavoura:1993su}.

Testing the theory is very difficult. To observe the righthanded
$W$-bosons seems  not possible, since the energy is too high.
However the $W$-bosons could result from a complex mixing among
the $W_L$ and $W_R$ gauge bosons, wherein the $W_R$ component is
extremely suppressed. In this case there will also be  a heavier
$W^\prime$-boson, wherein the $W_L$ component is suppressed. A
tiny mixing angle in the $W_L$-$W_R$ sector that could be detected
by precision experiments would immediately indicate a left-right
symmetric gauge sector. Similarly in an $SO(10)$-theory there will
be an extra neutral $Z^\prime$-boson, coupling to neutral currents
as investigated in ref.~\cite{ozer}. As a result we should expect
small deviations in the vector and axial-vector couplings of the
$Z$-boson which originate from the $SU(4)$ and $SU(2)_R$ gauge
fields. Another possibility seems to us to look for the proton
decay, which is expected around $10^{31-33}$ years.

\bibliographystyle{unsrt}
\bibliography{main}

\begin{thebibliography}{10}

\bibitem{Salam:1964ry}
A.~Salam and J.~C. Ward.
\newblock Electromagnetic and weak interactions.
\newblock {\em Phys. Lett.}, 13:168--171, 1964.

\bibitem{Weinberg:1967tq}
S.~Weinberg.
\newblock A model of leptons.
\newblock {\em Phys. Rev. Lett.}, 19:1264--1266, 1967.

\bibitem{Salam:1968rm}
A.~Salam.
\newblock Weak and electromagnetic interactions.
\newblock Originally printed in *Svartholm: Elementary Particle Theory,
  Proceedings Of The Nobel Symposium Held 1968 At Lerum, Sweden*, Stockholm
  1968, 367-377.

\bibitem{Glashow:1961tr}
S.~L. Glashow.
\newblock Partial symmetries of weak interactions.
\newblock {\em Nucl. Phys.}, 22:579--588, 1961.

\bibitem{Fritzsch:2002jv}
H.~Fritzsch and M.~Gell-Mann.
\newblock Current algebra: Quarks and what else?
\newblock 2002.

\bibitem{Fritzsch:1973pi}
H.~Fritzsch, Murray Gell-Mann, and H.~Leutwyler.
\newblock Advantages of the color octet gluon picture.
\newblock {\em Phys. Lett.}, B47:365--368, 1973.

\bibitem{Weinberg:1973un}
S.~Weinberg.
\newblock Nonabelian gauge theories of the strong interactions.
\newblock {\em Phys. Rev. Lett.}, 31:494--497, 1973.

\bibitem{Glashow:1970gm}
S.~L. Glashow, J.~Iliopoulos, and L.~Maiani.
\newblock Weak interactions with lepton - hadron symmetry.
\newblock {\em Phys. Rev.}, D2:1285--1292, 1970.

\bibitem{'tHooft:1972fi}
G.~'t~Hooft and M.~J.~G. Veltman.
\newblock Regularization and renormalization of gauge fields.
\newblock {\em Nucl. Phys.}, B44:189--213, 1972.

\bibitem{Aitchison}
Aitchison I.~J. R.
\newblock {\em Contemp. Phys}, 26:333, 1985.

\bibitem{Georgi:1974yf}
H.~Georgi, Helen~R. Quinn, and S.~Weinberg.
\newblock Hierarchy of interactions in unified gauge theories.
\newblock {\em Phys. Rev. Lett.}, 33:451--454, 1974.

\bibitem{groom:2000}
D.~E. Groom et~al.
\newblock Review of particle physics.
\newblock {\em Eur. Phys. J.}, C15:97, 2000.

\bibitem{Goldstone:1961eq}
J.~Goldstone.
\newblock Field theories with 'superconductor' solutions.
\newblock {\em Nuovo Cim.}, 19:154--164, 1961.

\bibitem{Goldstone:1962es}
J.~Goldstone, A.~Salam, and S.~Weinberg.
\newblock Broken symmetries.
\newblock {\em Phys. Rev.}, 127:965--970, 1962.

\bibitem{Higgs:1964pj}
Peter~W. Higgs.
\newblock Broken symmetries and the masses of gauge bosons.
\newblock {\em Phys. Rev. Lett.}, 13:508--509, 1964.

\bibitem{Higgs:1964ia}
Peter~W. Higgs.
\newblock Broken symmetries, massless particles and gauge fields.
\newblock {\em Phys. Lett.}, 12:132--133, 1964.

\bibitem{Higgs:1966ev}
Peter~W. Higgs.
\newblock Spontaneous symmetry breakdown without massless bosons.
\newblock {\em Phys. Rev.}, 145:1156--1163, 1966.

\bibitem{Weinberg:1971nd}
S.~Weinberg.
\newblock Mixing angle in renormalizable theories of weak and electromagnetic
  interactions.
\newblock {\em Phys. Rev.}, D5:1962--1967, 1972.

\bibitem{Weinberg:1979pi}
S.~Weinberg.
\newblock Conceptual foundations of the unified theory of weak and
  electromagnetic interactions.
\newblock {\em Rev. Mod. Phys.}, 52:515--523, 1980.

\bibitem{Politzer:1973fx}
H.~David Politzer.
\newblock Reliable perturbative results for strong interactions?
\newblock {\em Phys. Rev. Lett.}, 30:1346--1349, 1973.

\bibitem{Gross:1973id}
D.~J. Gross and F.~Wilczek.
\newblock Ultraviolet behavior of non-abelian gauge theories.
\newblock {\em Phys. Rev. Lett.}, 30:1343--1346, 1973.

\bibitem{Politzer:1974fr}
H.~David Politzer.
\newblock Asymptotic freedom: An approach to strong interactions.
\newblock {\em Phys. Rept.}, 14:129--180, 1974.

\bibitem{Jones:1974mm}
D.~R.~T. Jones.
\newblock Two loop diagrams in yang-mills theory.
\newblock {\em Nucl. Phys.}, B75:531, 1974.

\bibitem{Georgi:1974sy}
H.~Georgi and S.~L. Glashow.
\newblock Unity of all elementary particle forces.
\newblock {\em Phys. Rev. Lett.}, 32:438--441, 1974.

\bibitem{Georgi:1974my}
H.~Georgi.
\newblock The state of the art - gauge theories. (talk).
\newblock {\em AIP Conf. Proc.}, 23:575--582, 1975.

\bibitem{Gell-Mann:1976pg}
M.~Gell-Mann, P.~Ramond, and R.~Slansky.
\newblock Color embeddings, charge assignments, and proton stability in unified
  gauge theories.
\newblock {\em Rev. Mod. Phys.}, 50:721, 1978.

\bibitem{Machacek:1979tx}
M.~Machacek.
\newblock The decay modes of the proton.
\newblock {\em Nucl. Phys.}, B159:37, 1979.

\bibitem{Weinberg:1979sa}
S.~Weinberg.
\newblock Baryon and lepton nonconserving processes.
\newblock {\em Phys. Rev. Lett.}, 43:1566--1570, 1979.

\bibitem{Goldman:1979ij}
T.~J. Goldman and D.~A. Ross.
\newblock A new estimate of the proton lifetime.
\newblock {\em Phys. Lett.}, B84:208, 1979.

\bibitem{Ross:1980gh}
D.~A. Ross.
\newblock The calculation of the decay rate of the proton.
\newblock Presented at the Europhysics Study Conf., Erice, Italy, Mar 17-24,
  1980.

\bibitem{Goldman:1980ah}
T.~J. Goldman and D.~A. Ross.
\newblock How accurately can we estimate the proton lifetime in an su(5) grand
  unified model?
\newblock {\em Nucl. Phys.}, B171:273, 1980.

\bibitem{Goldhaber:1980dn}
M.~Goldhaber, P.~Langacker, and R.~Slansky.
\newblock Is the proton stable?
\newblock {\em Science}, 210:851--860, 1980.

\bibitem{Langacker:1992qb}
P.~Langacker.
\newblock Proton decay.
\newblock 1992.

\bibitem{kounnas}
D.~V. Nanopoulos K. A.~Olive C.~Kounnas, A.~Masiero.
\newblock {\em Grand Unification with and without Supersymmetry and
  Cosmological Implications}, pages 68--79,172--179.
\newblock Number~2 in International School for Advanced Studies Lecture Series.
  World Scientific Publishing Co. Pte. Ltd., P O Box 128, Farrer Road,
  Singapore 9128, 1983.

\bibitem{Langacker:1980js}
P.~Langacker.
\newblock Grand unified theories and proton decay.
\newblock {\em Phys. Rept.}, 72:185, 1981.

\bibitem{Fritzsch:1974nn}
H.~Fritzsch and P.~Minkowski.
\newblock Unified interactions of leptons and hadrons.
\newblock {\em Ann. Phys.}, 93:193--266, 1975.

\bibitem{ozer}
\"Ozer~A. D.
\newblock {\em $SO(10)$-Grand Unification and Fermion Masses}.
\newblock PhD thesis, Ludwig-Maximilians University, Arnold Sommerfeld Center,
  Theresienstr.37, DE-80333 M\"unchen, Prerintnumber: ASC-LMU 72/05, December
  2005.

\bibitem{Bauer:1952}
Friedrich~L. Bauer.
\newblock {\em Gruppentheoretische Untersuchungen zur Theorie der
  Spinwellengleichungen}, pages 153--159.
\newblock Number~13 in Matematisch-Naturwissenschaftliche Klasse 1952. Aus den
  Sitzungen der Bayerischen Akademie der Wissenschaften, Muenchen, 1952.

\bibitem{Pati:1974yy}
Jogesh~C. Pati and Abdus Salam.
\newblock Lepton number as the fourth color.
\newblock {\em Phys. Rev.}, D10:275--289, 1974.

\bibitem{Pati:1977jh}
Jogesh~C. Pati, Subhash Rajpoot, and Abdus Salam.
\newblock Complexions of neutral current phenomena within the left- right
  symmetric unified theory of quarks and leptons.
\newblock {\em Phys. Rev.}, D17:131--150, 1978.

\bibitem{Lavoura:1993su}
L.~Lavoura and L.~Wolfenstein.
\newblock Resuscitation of minimal so(10) grand unification.
\newblock {\em Phys. Rev.}, D48:264--269, 1993.

\end{thebibliography}

\appendix

\section{}
\noindent We have
\begin{equation}\label{scales}
\begin{split}
 M_R & = M_F  \exp \left[ \Delta_1 - \frac{\mathcal{ B}}{\mathcal{E} \mathcal{C} - \mathcal{F} \mathcal{B}   } \frac{1}{e^2} \left(\frac{3}{5} - \frac{8}{5}\sin^2\theta \right) + \frac{ \mathcal{E} }{ \mathcal{E} \mathcal{C}-\mathcal{F} \mathcal{B } } \left( \frac{1}{e^2}-\frac{8}{3}\frac{1}{g^2_5} \right) \right] \\
 M_C & = M_F \exp\left[  \Delta_2+ \frac{\mathcal{ C} - \mathcal{B} }{ \mathcal{E }\mathcal{C }- \mathcal{F }\mathcal{B } } \frac{1}{e^2} \left(\frac{3}{5} - \frac{8}{5}\sin^2\theta \right)   + \frac{  \mathcal{E }- \mathcal{F}   }{\mathcal{E }\mathcal{C} -\mathcal{F} \mathcal{B }  }  \left(\frac{1}{e^2}-\frac{8}{3}\frac{1}{g^2_5} \right) \right]\\
 M_G & = M_F \exp \left[ \Delta_3 +\frac{\mathcal{ C} - \mathcal{B} }{  \mathcal{E} \mathcal{C} - \mathcal{F} \mathcal{B}}  \frac{1}{e^2} \left(\frac{3}{5} - \frac{8}{5}\sin^2\theta \right)   + \frac{  \mathcal{E}- \mathcal{F }  }{ \mathcal{E} \mathcal{C }-\mathcal{F} \mathcal{B} } \left(\frac{1}{e^2}-\frac{8}{3}\frac{1}{g^2_5}\right)\right] \\
\end{split}
\end{equation}
The  $\Delta$'s appearing in these equations are composed of beta
functions and the ratio $\ln \frac{M_G}{M_C}$. They  are
explicitly found as
\begin{equation}
\begin{split}
 \Delta_1 & = \frac{\mathcal{D} \mathcal{B} - \mathcal{E} \mathcal{A} }{ \mathcal{ E} \mathcal{C}-\mathcal{F} \mathcal{B} } \ln \frac{M_G}{M_C}\\
 \Delta_2 & =  -\frac{  \mathcal{A} (\mathcal{E}-\mathcal{F}) + \mathcal{D }( \mathcal{C}-\mathcal{B})} {\mathcal{E }\mathcal{C} - \mathcal{F} \mathcal{B }   } \ln \frac{M_G}{M_C} \\
 \Delta_3 & = \left(1 - \frac{  \mathcal{A} (\mathcal{E}-\mathcal{F}) + \mathcal{D }(\mathcal{C} - \mathcal{B})} {\mathcal{E} \mathcal{C} - \mathcal{F }\mathcal{B}  }\right) \ln \frac{M_G}{M_C} \\
\end{split}
\end{equation}
The ratio is chosen by definition positive. If we substitute the
values of the beta functions, we find that
\begin{equation}
   \Delta_1 \geqslant 0 ,\ \ \ \ \ \  \Delta_2 \leqslant 0,\ \ \ \ \ \ \Delta_3 \leqslant 0
\end{equation}


\end{document}